\documentclass[12pt]{article}
\usepackage{amsmath}
\usepackage{graphicx}
\usepackage{enumerate}
\usepackage{natbib}
\usepackage{url} % not crucial - just used below for the URL 

\newcommand{\blind}{1}

\newtheorem{assumption}{Assumption}

\usepackage{amsmath, amsfonts, bbm, rotating, enumitem}
\usepackage{setspace}
\usepackage{authblk}   % For authors and affiliations
\usepackage{fullpage}
\usepackage{hyperref}
\usepackage{nameref}
\usepackage{verbatim}
\usepackage{natbib}
\usepackage{lmodern,xcolor}
\usepackage{graphicx}
\usepackage{amssymb}
\usepackage{epstopdf}

\usepackage{amsmath, amsfonts, bbm, rotating}
\usepackage{graphicx}
\usepackage{amssymb}
\usepackage{epstopdf}
\usepackage{bm}

\newcommand{\bOne}{\mathbbm 1}

\newcommand{\bx}{\bm x}

\newcommand{\bI}{\bm I}

\newcommand{\bOmega}{\bm \Omega}
\newcommand{\bomega}{\bm \omega}
\newcommand{\blambda}{\bm \lambda}

\newcommand{\bpi}{\bm \pi}
\newcommand{\bphi}{\bm \phi}

\newcommand{\logit}{\mbox{logit}}

\newcommand{\MALE}{\texttt{male}}
\newcommand{\TIME}{\texttt{time}}
\newcommand{\AGE}{\texttt{age}}
\usepackage{caption}
\usepackage{subcaption}
\usepackage{tikz}
\usetikzlibrary{shapes,decorations,arrows,calc,arrows.meta,fit,positioning,shadows}
\tikzset{
    every picture/.style=thick,
    -Latex,auto,node distance =1 cm and 1 cm,semithick,
    state/.style ={ellipse, draw, minimum width = 0.7 cm},
    point/.style = {circle, draw, inner sep=0.04cm,fill,node contents={}},
    bidirected/.style={Latex-Latex,dashed},
    el/.style = {inner sep=2pt, align=left, sloped}
}

\def\bSig\mathbf{\Sigma}

\def\bSig\mathbf{\Sigma}

\usepackage{graphicx} % Required for inserting images
\usepackage{booktabs}
\usepackage[tableposition=above]{caption}
\usepackage{pifont}

\begin{document}

\def\spacingset#1{\renewcommand{\baselinestretch}%
{#1}\small\normalsize} \spacingset{1}

%%%%%%%%%%%%%%%%%%%%%%%%%%%%%%%%%%%%%%%%%%%%%%%%%%%%%%%%%%%%%%%%%%%%%%%%%%%%%%

\if1\blind
{
  % Final version: Show authors and affiliations
  \title{\bf Hierarchical Latent Class Models for Mortality Surveillance Using Partially Verified Verbal Autopsies}
  
  % Author and affiliation block
  \author[1]{Yu Zhu}
  \author[1]{Zehang Richard Li}
  \affil[1]{Department of Statistics, University of California, Santa Cruz}

  \date{}  % No date displayed
  \maketitle
} \fi

\bigskip
\abstract{
Monitoring cause-of-death data is an important part of understanding disease burdens and effects of public health interventions. Verbal autopsy (VA) is a well-established method for gathering information about deaths outside of hospitals by conducting an interview to caregivers of a deceased person. It is usually the only tool for cause-of-death surveillance in low-resource settings. A critical limitation with current practices of VA analysis is that all algorithms require either domain knowledge about symptom-cause relationships or large labeled datasets for model training. Therefore, they cannot be easily adopted during public health emergencies when new diseases emerge with rapidly evolving epidemiological patterns. In this paper, we consider estimating the fraction of deaths due to an emerging disease. We develop a novel Bayesian framework using hierarchical latent class models to account for the informative cause-of-death verification process. Our model flexibly captures the joint distribution of symptoms and how they change over time in different sub-populations. We also propose structured priors to improve the precision of the cause-specific mortality estimates for small sub-populations. Our model is motivated by mortality surveillance of COVID-19 related deaths in low-resource settings. We apply our method to a dataset that includes suspected COVID-19 related deaths in Brazil in 2021. 
% We show that standard modeling approaches can be severely biased under selective verification and our model leads to more robust and accurate quantification of disease prevalence.
}

\noindent%
\vfill

\newpage
\spacingset{1.9} % DON'T change the spacing!

\section{Introduction}
Monitoring data describing cause of death is an essential component for understanding the burden of disease and evaluating public health interventions. Only about two-thirds of deaths worldwide are registered, and up to half of these deaths do not have an assigned cause, with the least information available from countries with the most need \citep{world2021civil}. A widely used tool to obtain information on causes of death when a medically certified cause-of-death is not available is verbal autopsies (VA). VA involves a structured questionnaire administered to family members or caregivers of a recently deceased person. The VA interviews collect information about the circumstances, signs, and symptoms leading up to death. VAs are widely adopted both in research settings and by national statistical offices \citep[e.g., ][]{maher2010translating,nkengasong2020improving}. Such information is either analyzed by physicians or with statistical algorithms to assign causes of death.

During public health emergencies when a new disease emerges, VA is usually the only feasible tool to gather information on causes of death in many low-resource settings, e.g., during the Ebola hemorrhagic fever outbreaks \citep{alert2003investigating}, the Dengue epidemic \citep{saqib2014retrospective}, and the COVID-19 pandemic \citep{de2020validation,  rosen2021development}. In such settings, physician review of VAs is usually infeasible since the process is time-consuming and expensive, especially during a public health crisis, and cause-of-death assignment needs to be performed by automatic VA algorithms in any medium- to large-scale mortality surveillance settings.
While significant advances have been made in automatic cause-of-death assignment algorithms in the last several decades \citep[see e.g., ][for a review]{chandramohan2021estimating}, using VAs for the purpose of monitoring mortality patterns and trends due to new diseases remains challenging and largely unexplored in the existing literature. The lack of guidance on using algorithms to analyze VA for rapid mortality surveillance presents a critical gap for governments and public health researchers, as evidenced by the latest COVID-19 pandemic. 

In order to be prepared for future outbreaks of new diseases, two critical challenges need to be addressed with VA algorithms. 
First of all, all existing VA algorithms operate with a pre-defined set of symptoms and causes and their relationships are either provided by physicians and assumed to be known \citep{byass2019integrated, mccormick2016probabilistic} or need to be estimated from high-quality training data, i.e., reference deaths with known causes verified through a separate mechanism \citep{tsuyoshi2017, moran2021bayesian}. In the case of an emerging disease, the relationship between symptoms and the new cause of death is generally unknown and needs to be estimated from the limited training data during the outbreak. Unlike traditional settings, the reference deaths with verified causes during an outbreak are usually not a random sample of the population due to logistical constraints and changing public health priorities. Such reference deaths can exhibit different symptom-cause relationships, and VA algorithms trained on these deaths could lead to biased estimates. This issue is similar to the verification bias in the diagnostic test literature, where the accuracy of a diagnostic test can be distorted by evaluation based on the patients with verified disease status only \citep{zhou1998correcting}. No guidelines exist on when and how to account for the selection process of reference death in VA research.

In addition, unlike routine analysis of VA with a single and static population, it is often of interest to monitor how the fraction of deaths due to each disease changes over time and how it varies across fine-scale demographic groups. During the course of an outbreak, the population mortality profiles can change rapidly due to both disease dynamics and changing treatment and intervention measures. Thus, the accuracy of cause-of-death assignment methods trained on data from the early phase of an outbreak may deteriorate over time. It is important to flexibly adapt the model to the changing distribution over time and across related sub-populations. A related line of work by \citet{moran2021bayesian} and \citet{kunihama2024bayesian} shows that modeling covariate-dependent symptom distributions can improve VA algorithms in general. However, their target of inference is still the overall cause-specific mortality fraction in a static population, and covariates are introduced only to improve the estimation of the joint distribution of symptoms. Directly extending single-population models by treating the sub-populations as independent can lead to inefficient estimates when sub-population sample sizes are small.

In this paper, we address these challenges through the lens of distribution shift. We show that the issue with emerging disease monitoring using VAs can be viewed as a lack of generalizability of algorithms from a non-representative training dataset to target populations \citep{li2021bayesian, wu2021tree}. We develop a unified hierarchical latent class modeling framework informed by the causal structure of the cause-of-death verification process. Our model allows direct estimation of the fraction of deaths due to the disease across multiple sub-populations over time. We also propose a novel set of structured priors to efficiently borrow information and improve the precision of estimates.

The rest of the paper is organized as follows. Section \ref{sec:data} introduces a COVID-19 surveillance dataset from Brazil that motivates this study. Section \ref{sec:model} introduces our proposed hierarchical latent class model with structured priors over sub-populations. Section \ref{sec:mcmc} develops an efficient Markov Chain Monte Carlo (MCMC) algorithm for posterior sampling. Section \ref{sec:result} demonstrates the proposed model with a series of analyses using both simulated data and the COVID-19 surveillance dataset from Brazil. Section \ref{sec:discuss} concludes with a discussion of limitations and future work on using VA for mortality surveillance.

\section{Brazil COVID-19 surveillance data} \label{sec:data}
\begin{figure}[!tb]
    \centering
    \includegraphics[width = 0.8\textwidth]{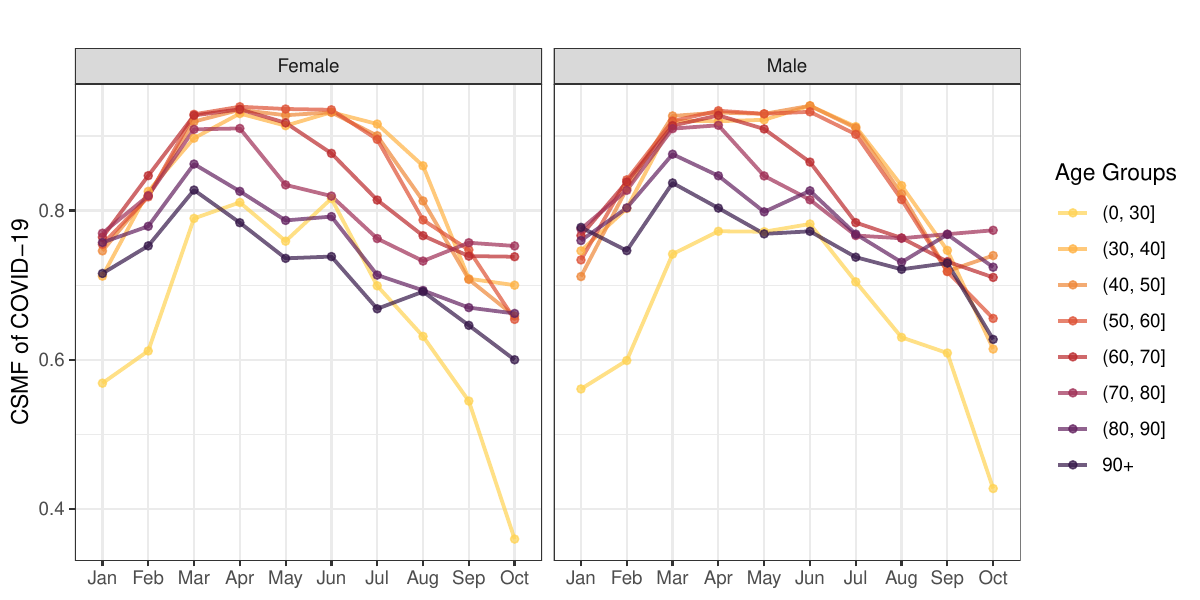}
    \caption{Fraction of deaths due to COVID-19 among the deaths suspected of COVID-19 across different age groups and sexes over time.}
    \label{fig:true_prevalence}
\end{figure}

\begin{figure}[!tb]
    \centering
    \includegraphics[width = .8\textwidth]{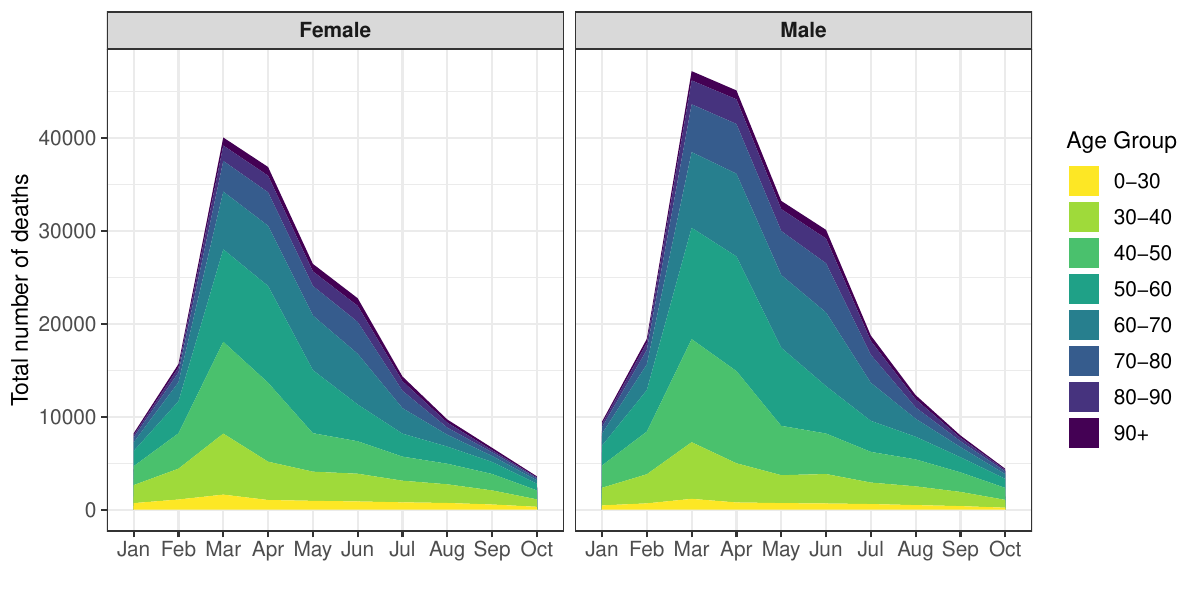}
    \caption{Total number of deaths in the COVID-19 surveillance dataset by age group, sex, and month.}
    \label{fig:sample_size}
\end{figure}

 Our study is motivated by the task of identifying COVID-19 deaths in a flu syndrome surveillance dataset collected in Brazil from January to October 2021. All deaths in this database had severe acute respiratory syndrome by COVID, other respiratory diseases or chronic diseases. The dataset contains the independently assigned underlying cause of death for $411,491$ deaths. The dataset consists of $p = 14$ binary indicators for each death, including symptoms such as fever, vomiting, loss of taste and smell, etc. Additional information on vaccine, PCR and antigen test results are also recorded but they are largely missing and thus are excluded from our analysis. Within this population of deaths suspected of COVID-19, the proportion of deaths with the final classification of being COVID-19 related increases from January to April and declines afterward. We consider eight age groups, where the first includes deaths between 0 and 30 years old and the subsequent groups are defined by ten-year increments. We let the first age group span 30 years because deaths under $20$ years old are rare in this population. Figure \ref{fig:true_prevalence} shows the true fraction of deaths due to COVID-19 among these suspected deaths, within each sub-population. Figure \ref{fig:sample_size} shows that the sample sizes across sub-populations are notably imbalanced, with particularly small samples observed in January and October, as well as in the first and last age groups.

% This is an example of a new parapgraph with a numbered footnote\footnote{\url{https://data.gov.uk/}} and a second footnote marker.\footnote{Example of footnote text.}

\section{Methods}\label{sec:model}
\subsection{Assumptions on verification mechanism}\label{sec:stratification}
Let $X_i \in \{0, 1\}^p$ denote the $p$-dimensional vector of signs/symptoms for the $i$-th death and $Y_i \in \{0, 1\}$ denote whether the underlying cause of death is COVID-19 related. The overall fraction of deaths due to COVID-19, or the cause-specific mortality fraction (CSMF) of COVID-19, is then $p(Y)$. In this paper, we focus on symptom surveys designed specifically to identify COVID-19 deaths. As a result, there is limited information available to distinguish among non-COVID-19 causes. Therefore, we treat $Y$ as a binary variable.
% Consider a dataset $\TT$ and assume that it is a simple random sample of the target population. The key targets of inference for VA analysis are usually the population-level cause-specific mortality fraction (CSMF), $p_{\TT}(Y)$, and the individual-level cause-of-death assignment rules $p_{\TT}(Y|X = x)$. For mortality surveillance of emerging diseases, we are also interested in these quantities stratified by different sub-populations of interest, e.g., time periods and age groups. 
% To simplify notation, we assume that $\TT$ itself is the population of interest, rather than a random sample of the population. This is the standard assumption made by all existing literature, but we will demonstrate in later sections that we can relax this assumption in certain situations and estimate the above probabilities for a related but different population. 
% We assume that there is a training dataset $\TT_{1} \in \TT$ with known cause of death assignment, verified through an independent mechanism. We denote the unlabeled deaths $\TT_{0} = \TT \setminus \TT_{1}$. 
% This is especially common in the context of emerging diseases, where resources may only allow a subset of deaths to be verified within the same population. 

We focus on the situations where the cause of death was known for only a subset of deaths. Typically, these causes are assigned and verified by physicians based on extensive reviews of all available information for each death. Let $L_i \in \{0, 1\}$ be the indicator of whether the cause of death for the $i$-th death is known. 
% While one may treat the labeled and unlabeled deaths as forming independent domains and apply methods proposed in \citet{li2021bayesian} and \citet{wu2021tree}, we show in this section that such separation may be unnecessary under certain verification mechanisms. 
% In the rest of this section, we discuss and establish conditions under which different models can be used or adapted to provide valid inference for the estimands of interest. 
% As an overview, Figure \ref{fig:dag:simple} shows the directed acyclic graph (DAG) corresponding to the factorization of the joint distribution of $(X, Y, S)$ underlying different selection mechanisms. It is worth noting that in the existing VA literature, the causal pathway between the collected signs and symptoms and causes of death has been largely ignored. While most of the indicators representing medical symptoms can be considered as caused by the underlying disease, demographic indicators and existence of risk factors cannot be treated as causal descendants of the underlying disease. Such distinctions becomes important when we analyze the effect of verification bias, as it leads to different estimators for the population prevalence as demonstrated in this section. 
 % that while we consider an arrow pointing from $Y$ to $X$ in all of the DAGs, it does not encode a causal relationship between $X$ and $Y$ as $X$ is complex and $X$usually contain indicators that both before (e.g., whether the decedent had close contact with a confirmed COVID case) and after (e.g., whether the decedent had fever) the underlying disease.
Existing literature typically assumes that $L_i = 1$ for all deaths with VAs in some training dataset and $L_i = 0$ for all deaths in the target dataset. When these two datasets arise from different populations, modeling assumptions on the two data distributions are necessary in order to ensure transportability of the classifier built on the training data \citep{li2021bayesian}.

In the context of mortality monitoring during outbreaks, the deaths with verified causes are usually a subset of the target population.  It is crucial to understand how the deaths are selected into the training dataset. If the training subset is not a simple random sample of the population, models estimated using the verified deaths may be inadequate to generalize to the entire population directly. In surveillance settings, it is common to over-sample deaths of certain characteristics, e.g., deaths in certain age groups or from certain locations, in order to increase the chance of identifying certain types of deaths, or simply due to logistical reasons. The selection process may also change over time. We refer to any such process that generates $L$ as the verification mechanism. 

Valid inference of the CSMF is possible if the verification mechanism is conditionally ignorable. Assumption \ref{ass:strat} formally characterizes this condition. 

\begin{assumption}\label{ass:strat}
The selection probability of receiving a verified cause of death only depends on known stratification variables $D$ and symptoms $X$, but not the cause of death $Y$ or other unobserved variables associated with $Y$. That is, $L \perp Y \mid X, D$. 
\end{assumption}

Assumption \ref{ass:strat} is equivalent to assuming that the cause of death is missing at random \citep{rubin1976inference}. 
The choice of $D$ should be informed by how the reference deaths with verified causes are selected. In many situations with mortality surveillance, the selection mechanism is known and depends on observable quantities. When it is not the case, the choice of $D$ is contextual. Including more covariates in $D$ mitigates the bias from potential violation of the assumption but increases model complexity. Proxy variables such as time periods and region indicators may be included in $D$ to mitigate the effect due to unobserved variables that vary over time or space.

Figure \ref{fig:dag:D} illustrates the data generating process of random variables under Assumption \ref{ass:strat}, where no arrows point from $Y$ to $L$ directly and there are no unobserved confounders. It is worth noting that in the data generating process, we assume the observed variables $X$ are generated conditional on the underlying cause of death $Y$, thus the arrow from $Y$ to $X$. This assumption follows the convention in VA analysis where symptoms are treated as consequences of the underlying disease \citep{king2008verbal, li2021bayesian}. Prediction tasks with this type of structure are usually referred to as being anticausal \citep{scholkopf2012causal}. 

The anticausal structure can be further refined in the context of VA. Conceptually, we may partition the variables collected from VA into $X = (X_C, X_E)$ where $X_C$ are variables that affect the risk of the cause of death, e.g., demographic variables, and $X_E$ are the variables affected by the cause of death, e.g., medical symptoms. This leads to the data generating process with a mixed causal structure shown in Figure \ref{fig:dag:DXc}. Our target of inference in this setting is the sub-population CSMFs, $p(Y \mid D, X_C)$, instead of the overall CSMF, $p(Y)$. Previous VA models have also explored separating demographic variables from medical symptoms, and such separation of covariates and symptoms has been shown to improve cause-of-death classification \citep{moran2021bayesian, kunihama2024bayesian}. However, their focus was on improving the estimation of the overall CSMF, and the covariates are still treated as downstream consequences of the cause of death in the assumed model structures. In contrast, our model explicitly distinguishes between upstream covariates and downstream symptoms, and explicitly targets covariate-dependent CSMFs. To simplify notation in the rest of the paper, we note that $X_C$ and $D$ play the same role in terms of the model, and differ only in whether they are collected from the VA survey or supplied as external information. Thus we omit $X_C$ and $X_E$, and use $D \in {1, ..., G}$ to denote the combination of all levels of variables defining stratification, and $X$ to denote the rest of the indicators from the VA survey, as in Figure \ref{fig:dag:D}.

\begin{figure}[tb]
\centering
\begin{subfigure}[t]{0.48\textwidth}
  \begin{center}
  \resizebox{.7\textwidth}{!}{
      \begin{tikzpicture}
       \node (X) at (0,0) {$X$};
       \node (Y) at (-2,0) {$Y$};
       \node (S) at (2,0) {$L$};
       \node (D) at (0,1.5) {$D$};
       % \node (W) at (-2,1.5) {$W$};
       \path (Y) edge (X);
       \path (X) edge (S);
       \path (D) edge (X);\path (D) edge (Y);\path (D) edge (S);
       % \path (W) edge (Y);\path (W) edge (X);
       \end{tikzpicture}
    }
    \end{center}
\caption{Ignorable verification given stratum $D$ and symptoms $X$.}
\label{fig:dag:D}
\end{subfigure}
\hfill
\begin{subfigure}[t]{0.48\textwidth}
  \begin{center}
  \resizebox{.7\textwidth}{!}{
      \begin{tikzpicture}
       \node(D) at (2,2) {$D$};
       \node(XC) at (0,2) {$X_C$};
       \node(Y) at (0,0) {$Y$};
        \node(XE) at (2,0) {$X_E$};
        \node(S) at (4,0) {$L$};
        \path (XC) edge (Y);
        \path (Y) edge (XE);
        \path (XE) edge (S);
        \path (D) edge (Y);
        \path (D) edge (XE);
        \path (D) edge (S);
        \path[bend right=0] (D) edge (XC);
        \path[bend right=0] (XC) edge (XE);
        \path[bend right=0] (XC) edge (S);
       \end{tikzpicture}
    }
    \end{center}
\caption{Ignorable verification given stratum $D$, covariate $X_C$ and symptoms $X_E$.}
\label{fig:dag:DXc}
\end{subfigure}
\caption{Directed acyclic graph (DAG) describing different data generating processes and sample selection mechanisms. $D$ and $X_C$ are observed for all deaths and $Y$ is only observed when $L = 1$.}
\label{fig:dag}
\end{figure}

% \subsection{Semi-supervised approach}\label{sec:semi}

% \begin{align}\nonumber
% \hat p_{\TT}(Y) &= \sum_{x, d} \hat p(Y | X = x, D = d)p(X = x, D = d) \\ \nonumber
%    % &= \sum_x \hat p(Y | X = x, D = d)p(X = x, D = d) \\\nonumber
%    &= \sum_{x, d} \hat p(Y | X = x, D = d, S = 1)p(X = x, D = d) \\ 
%    &\approx \frac{1}{n} \sum_{i = 1}^n \hat p_{\TT_1}(Y | X = x_i, D = d_i)
% \label{eq:simple}
% \end{align} 
% A major drawback of such discriminant classification models, however, is that they do not take into account the potentially large collection of unlabeled VA data.
% On the other hand, semi-supervised approaches that model both the labeled and unlabeled data can achieve better predictive performance in tasks where $Y$ is the `cause' of the observed covariates $X$, as the distribution of $X$ in the unlabeled data also contains information about the conditional distribution of $X$ given $Y$ \citep{kugelgen2020semi}. Classification tasks with this type of structure are usually referred to as being `anticausal' \citep{scholkopf2012causal}. The VA problem is anticausal in nature because the majority of the symptoms and indicators are consequences of the underlying cause of death \citep{king2008verbal, li2021bayesian}. Figure \ref{fig:dag:D} illustrates the data generating process of the observable quantities under the assumed anticausal structure and the conditional independence relationship in Assumption \ref{ass:strat}. Therefore, in this work, we treat the unknown cause of death $Y$ as missing data and fit our models on all the deaths together.

\subsection{Hierarchical latent class model}\label{sec:latent}
We now turn to model specification. 
Given a choice of $D$ that renders the selection mechanism conditionally ignorable, one may learn a binary classification rule using the training dataset. The overall CSMF can be estimated by  $\hat p(Y) = \frac{1}{n} \sum_{i = 1}^n \hat p(Y \mid X = x_i, D = d_i, L = 1)$. Sub-population CSMFs $p(Y \mid D = d)$ can also be obtained by aggregation in a similar fashion. A major drawback of such discriminant classification models is that they do not take into account the large collection of unlabeled VA data, which provides rich information on the distribution of $X$ in anticausal problems \citep{kugelgen2020semi}. Therefore, we consider a semi-supervised approach that models both labeled and unlabeled data together. 

\begin{figure}[!h]
    \centering
    \includegraphics[width = \textwidth]{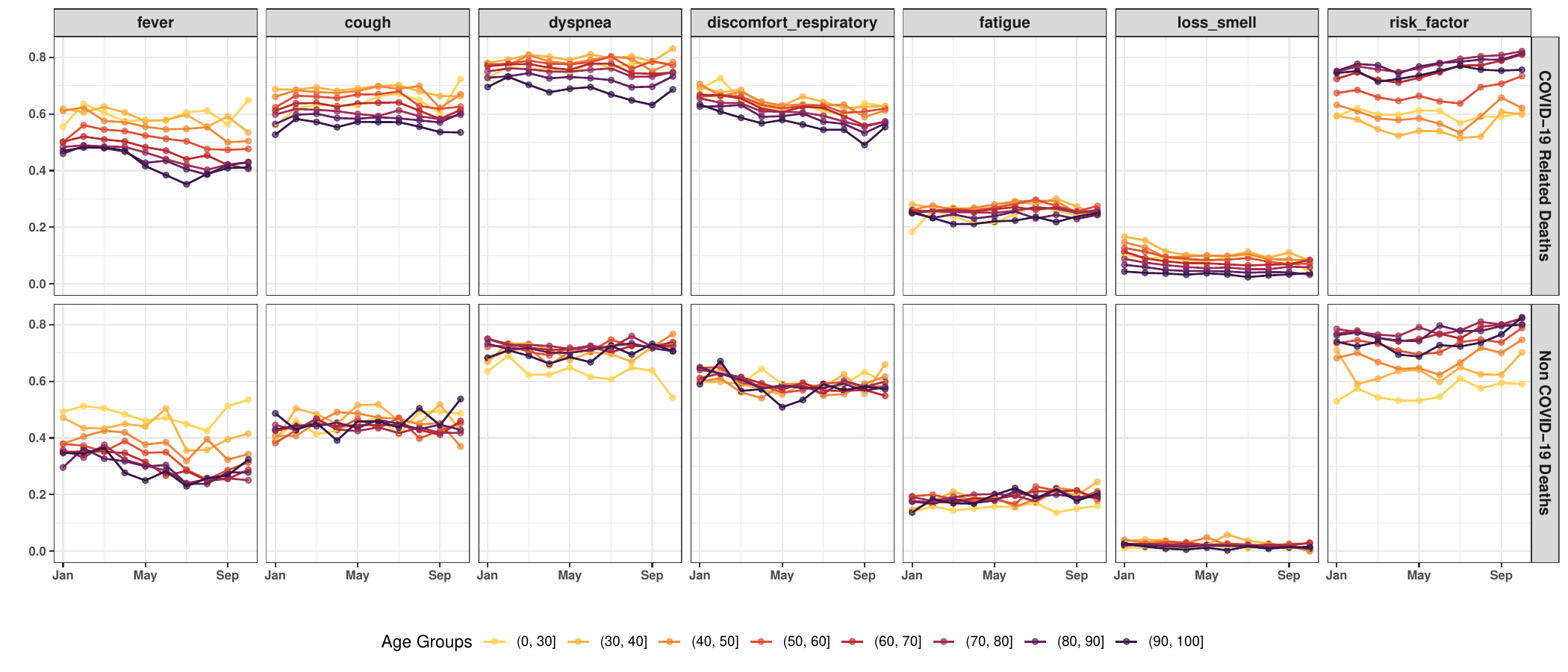}
        \caption{Time-varying proportion of deaths reporting different symptoms across eight age groups and two causes of death, among deaths the Brazil VA dataset.}
    \label{fig:symptoms_dist}
\end{figure}

\begin{figure}[!h]
    \centering
    \includegraphics[width = 0.7\textwidth]{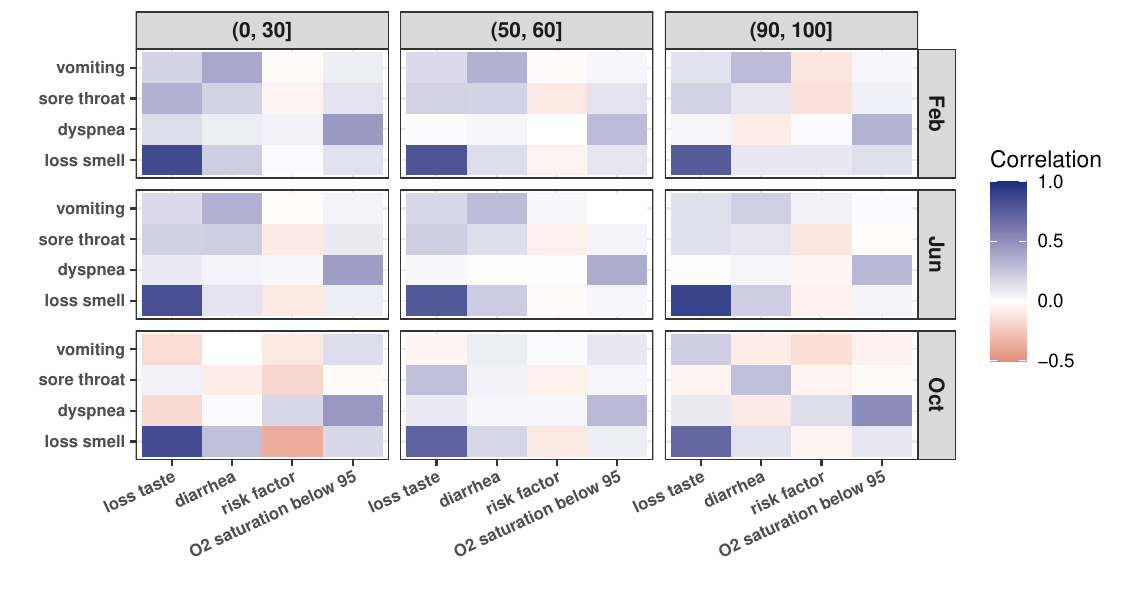}
    \caption{Matthews correlation coefficient \citep{Matthews1975} of a subset of symptoms among deaths related to COVID-19 among deaths in three age groups from three time periods in the Brazil VA dataset. Missing data are removed when calculating the correlation coefficient. Results for all age groups and time periods with bootstrap uncertainty intervals are reported in the Supplementary Materials.}
    \label{fig:correlation_heatmap}
\end{figure}

Our assumed data generating process in Figure \ref{fig:dag} leads to the factorization of the joint distribution $p(X, Y, D) = p(D)p(Y\mid D)p(X \mid Y, D)$. Many existing VA models simplify the last conditional probability into $p(X \mid Y)$ by assuming the conditional distribution of symptoms given causes is transferable \citep[e.g., ][]{mccormick2016probabilistic, byass2019integrated}. Such assumptions are usually violated when considering symptom distributions across deaths of different sexes and age groups \citep{moran2021bayesian} or over different populations \citep{li2021bayesian}. To illustrate the conditional dependence on $D$, we use the COVID-19 dataset described in Section \ref{sec:data} to compute the empirical symptom prevalence and the Matthews correlation coefficient \citep{Matthews1975} of some selected symptoms across different age groups and over time. Figure \ref{fig:symptoms_dist} shows that the symptom prevalence varies across both time and age groups. For example, among COVID-19 related deaths, the proportion of observations with fever as a symptom decreases over time and as age increases. Figure \ref{fig:correlation_heatmap} shows that the dependence among symptoms also varies significantly with the age and time period of the death. 
% For instance, the correlation between the symptoms `loss of taste' and `vomiting' under the younger age group of $0-30$ between February and October changes from positive to negative, which could be influenced by temporal and environmental factors. This suggests that certain signs/symptoms can become more or less informative over different sub-populations and over time for cause-of-death assignment and prevalence quantification. 

In order to capture the heterogeneity of symptoms distributions over different sub-populations in a parsimonious way, we adopt the nested latent class model framework developed in \citet{li2021bayesian} and \citet{wu2021tree}, where individual-level latent class membership indicator $Z_i \in \{1, 2, ..., K\}$ is introduced to flexibly capture the dependence of $X$. We assume the following data generating process,
\begin{align}
Y_i  \mid D_i = g&\sim \mbox{Bern}(\pi^{(g)}),\\
Z_i \mid Y_i = c, D_i = g&\sim \mbox{Cat}(\blambda_{c}^{(g)}), \\
X_{ij} \mid Y_i  = c, Z_i = k &\sim \mbox{Bern}(\phi_{ckj}), \;\;  j = 1, ..., p.
\end{align} 
The introduction of the latent class nested under each cause of death leads to a flexible characterization of cause-specific dependent symptom profiles that vary across strata. That is, after integrating out the latent indicators, the conditional distribution of symptoms given a cause of death and sub-population is $p(X_i \mid Y_i = c, D_i = g) = \sum_k \lambda_{ck}^{(g)} \prod_{j} \phi_{ckj}^{x_{ij}} (1 - \phi_{ckj})^{1 - x_{ij}}$. When $K = 1$, the model reduces to the commonly used conditional independent model in \citet{mccormick2016probabilistic}. In order to facilitate easier interpretations and mitigate the risk of overfitting, we fix the latent response probabilities $\bphi$ to be invariant across strata and let the sub-populations mix differently over the latent symptom profiles. With sufficiently many latent classes $K$, this representation is flexible enough to capture any multivariate discrete distribution \citep{dunson2009nonparametric}. 

More generally, $K$ can be treated as random and inferred from data, as is usually carried out in the Bayesian nonparametric literature \citep{hjort10, hannah2011dirichlet}.
In this work, we fix $K = 10$ to simplify the computation. Sensitivity analysis for different choices of $K$ can be found in the Supplementary Materials. For the analysis in this paper, where there are only $p = 14$ symptoms, $K = 10$ is sufficiently large, and posterior inference of the CSMF does not change much when $K$ is further increased.

The prior of $\pi^{(g)}$ needs to be designed for the specific choice of stratification variable $D$. We discuss our structured prior for $\pi^{(g)}$ in the next subsection. For the rest of the latent parameters, we put stick-breaking priors on $\blambda_{c}^{(g)}$ and conjugate Beta priors on $\bphi$, i.e.,
\begin{align*}
  % \lambda_{c1}^{(g)} &= V_{c1}^{(g)}, 
  \lambda_{ck}^{(g)} &= V_{ck}^{(g)} \prod_{l < k} (1 - V_{cl}^{(g)}), \;\;\;
 V_{ck}^{(g)}  \sim \mbox{Beta}(1, \omega_{c}^{(g)}), \text{ for } k = 2, ..., K, \\ 
 V_{cK}^{(g)} &= 1, \;\;\; \omega_c^{(g)} \sim \mbox{Gamma}(a_\omega, b_\omega), \;\;\;
\phi_{ckj} \sim\mbox{Beta}(a_{\phi}, b_{\phi}).
\end{align*}

We treat any unknown $Y_i$ as missing data without modeling the verification process due to Assumption \ref{ass:strat}. Missing values in the symptoms are also common in practice. We follow the practice of current VA algorithms by assuming they are also missing at random \citep{tsuyoshi2017}. 

\subsection{Stratification and structured priors}\label{sec:structure}

A main challenge when considering fine stratification of a population is that inevitably some sub-populations will contain only a small number of deaths.  Previous work involving cause-of-death assignment in multiple populations all assume independent priors for $\pi^{(g)}$ \citep[see e.g.,][]{mccormick2016probabilistic, li2021bayesian, wu2021tree}. For mortality monitoring of small sub-populations, however, it is usually more appropriate to borrow information across related sub-populations in order to improve the stability and interpretability of estimated CSMFs. Similar ideas have been extensively studied for survey data in the context of small area estimation \citep[e.g.,][]{rao2015small, gao2021improving}. Small area estimation methods have not been explored in VA analysis. Unlike the traditional small area estimation literature, the outcome of interest, i.e., the cause of death, cannot be directly collected and needs to be estimated instead. However, the key idea of leveraging the structured relationship of sub-populations and borrowing information across strata applies naturally to mortality surveillance using VAs.    

To fix notation, here we describe the model where sub-populations $D$ are defined by three stratification variables. Let $S_i \in \{1, 2\}$ indicate sex (1 = female and 2 = male), $T_i \in \{1, 2, ..., T\}$ indicate the time period, and $A_i = \{1, 2, ..., A\}$ indicate the age group of the $i$-th death respectively. The hierarchical latent class model with $D_i = (S_i, T_i, A_i)$ can be written as 
\begin{align}
    p(Y_{i} = 1 \mid D_i) &= \pi_{s[i], t[i], a[i]},\\
    p(Z_{i} = k \mid Y_{i} = c, D_i) &= \lambda_{ck}^{(s[i], t[i],a[i])}, \;\;\;\; c = 0, 1.
\end{align}
We can then encode different prior beliefs into the prior distributions for $\bpi$. In this paper, we consider a linear model on the logit scale,
\begin{equation}
\pi_{sta} = \logit^{-1}(\mu + \alpha^{\MALE}\bOne_{s = 1} + \alpha^{\TIME}_{t} + \alpha^{\AGE}_{a} + \epsilon_{sta}), 
\end{equation}
% \red{
% Let's have one more model extending the additive form with
% \[
% \pi_{ta} = \logit^{-1}(\alpha_a^{\AGE} + \gamma_{ta} + \epsilon_{ta}), \;\;\;\;
% (\gamma_{1a}, ..., \gamma_{Ta}) \sim RW(\sigma^2) \; \mbox{for }  a = 1, ..., A
% \]
% and $\sum_{t} \gamma_{ta} = 0$ for $a = 1, ..., A$, and $\alpha_a^{\AGE} \sim N(0, 100)$ no need for another intercept if all age groups have a fixed effect. By the way, since we are only doing time and age, maybe denote $\alpha^\TIME$ and $\alpha^\AGE$ with just $\alpha$ and $\beta$ instead. Another thing to consider adding here (I think it will save us a lot of trouble justifying) is to have an indicator of sex in all models. So for example in this new model, 
% \[
% \pi_{tas} = \logit^{-1}(\eta \bm 1_{s = 1} + \alpha_a^{\AGE} + \gamma_{ta} + \epsilon_{ta}), 
% \]
% where $s = 0, 1$ is the binary indicator of sex. Since we know distribution given male/female do not change much, let us use a single offset here to start with.
% }
where $\bOne_{s = 1}$ denotes the indicator function that takes value of $1$ when $s = 1$ and $0$ otherwise. We model $\mu$ and $\alpha^{\MALE}$ as fixed effects with independent $N(0, 100)$ priors.  We let $\epsilon_{sta} \overset{iid}{\sim} N(0, \sigma_\epsilon^2)$ be an unstructured interaction term that captures the deviation from the main additive decomposition. For the time and age effects, We consider the following three prior specifications:

\begin{enumerate}
    \item {Fixed effect:} $\alpha^{\TIME}_{t} \sim N(0, 100)$ and $\alpha^{\AGE}_{a} \sim N(0, 100)$.
    \item {Independent random effect: }
    $\alpha^{\TIME}_{t}  \sim N(0, \sigma_{\TIME}^2)$ and 
    % $\sigma_{\TIME}^2 \sim \mbox{Inv-Gamma}(0.5, 0.5)$ \\
    $\alpha^{\AGE}_{a}  \sim N(0, \sigma_{\AGE}^2)$,
    % $\sigma_{\AGE}^2 \sim \mbox{Inv-Gamma}(0.5, 0.5)$.
    \item {First order random walk (RW1): } 
    % $\alpha^{\TIME}_{1} \sim N(0, 100)$,  $\alpha^{\AGE}_{1} \sim N(0, 100)$\\
    $\alpha^{\TIME}_{t} \mid \alpha^{\TIME}_{t-1} \sim N(\alpha^{\TIME}_{t-1}, \sigma_{\TIME}^2)$  \, and $\alpha^{\AGE}_{a} \mid \alpha^{\AGE}_{a-1}\sim N(\alpha^{\AGE}_{a-1}, \sigma_{\AGE}^2)$ for $t = 2, ..., T$ and $a = 2, ..., A$. $\alpha^{\TIME}_{1} \sim N(0, 100)$ and $\alpha^{\AGE}_{1} \sim N(0, 100)$.
\end{enumerate}

For the hyperpriors, we use $\mbox{Inv-Gamma}(0.5, 0.0015)$ prior for the variance parameters of independent random effects and $\mbox{Inv-Gamma}(0.5, 0.0009)$ prior for the random walk models. These prior choices lead to a 95\% prior interval of $[0.5, 2]$ for the residual odds ratio \citep{fong2010bayesian}. The prior for the RW1 model is different because the variance parameter is conditional rather than marginal. Similar weakly informative priors have been used in small area estimation models with structured priors \citep[e.g., ][]{mercer2015space}. The hyperpriors need to be chosen carefully in practice to avoid over- and under-smoothing. For the analysis presented in this paper, the results are not sensitive to the choice of hyperpriors, as long as they are only weakly informative. Prior sensitivity analysis can be found in the Supplementary Materials.

The three prior specifications differ in the amount of information shared across strata. In the fixed effect model, no information is shared except for the additive structure of the time and age effect. The independent random effect model further shrinks these effects to the common mean. The random walk model shrinks the first order difference in these effects so that strata with similar ages or time periods will have similar CSMFs.

% Posterior sampling of all four models can be performed using Gibbs sampling with P\'{o}lya-Gamma augmentations for the three structured priors.

% \red{Here is potentially where we describe what structured priors we use for the latent class mixing weights. I think we can just pick one parameterization, e.g., random walk with additive structures (or something else that works well compared to data), and not vary and experiment multiple options, since there is no direct observations guiding us what model to use anyway. But we can compare the structured $\lambda$ with the current independent stick breaking prior we use (maybe only in a subset of the experiments below).}

\section{Posterior inference} \label{sec:mcmc}
The posterior distribution of the model parameters is not available in closed form, but we can easily obtain posterior samples from a Gibbs sampler as follows. 

\begin{enumerate}
    \item Sample $Y_i \mid \bm{X_i}, \bpi, \bm{\phi}, \bm{\lambda}$ for $i$ where $L_i = 0$ with
            % \[
            %   p(Y_i = 1 \mid \bm{X_i}, \bpi, \bm{\phi}, \bm{\lambda}) = \frac{\pi^{(g[i])} \sum_{k=1}^K \lambda_{1k}^{(g[i])} \prod_{j=1}^p  \phi_{1kj}^{X_{ij}}(1-\phi_{1kj})^{1-X_{ij}}}{
            %   \sum_{c=0}^{1} (\pi^{(g[i])})^c(1-\pi^{(g[i])})^{1-c}  \sum_{k=1}^K \lambda_{ck}^{(g[i])} \prod_{j=1}^p \phi_{ckj}^{X_{ij}}(1-\phi_{ckj})^{1-X_{ij}}}
            % \]
   \begin{align*}
                p(Y_i = c \mid \bm{X_i}, \bpi, \bm{\phi}, \bm{\lambda}) \;\propto \;
                &\pi_{s[i], t[i], a[i]}^c
                (1-\pi_{s[i], t[i], a[i]})^{1-c} \\
                &\times \sum_{k=1}^K \lambda_{ck}^{(s[i], t[i], a[i])} \prod_{j=1}^p  \phi_{ckj}^{X_{ij}}(1-\phi_{ckj})^{1-X_{ij}}.
    \end{align*}
   % where g[i] = [s[i], t[i], a[i]].
  \item Sample $Z_i \mid Y_i = c, \bm{X_i}, \bm{\phi}, \bm{\lambda}$ for $i = 1, ..., n$ with 
            % \[
            %   p(Z_i = k \mid Y_i = c, \bm{X_i}, \bm{\phi}, \bm{\lambda}) = \frac{\lambda_{ck}^{(g[i])} \prod_{j=1}^p\phi_{ckj}^{X_{ij}}(1-\phi_{ckj})^{1-X_{ij}}}{
            %   \sum_{k = 1}^{K}\lambda_{ck}^{(g[i])} \prod_{j=1}^p \phi_{ckj}^{X_{ij}}(1-\phi_{ckj})^{1-X_{ij}}
            %   }
            % \]    
   \[
                p(Z_i = k \mid Y_i = c, \bm{X_i}, \bm{\phi}, \bm{\lambda}) \propto \lambda_{ck}^{([s[i], t[i], a[i])} \prod_{j=1}^p\phi_{ckj}^{X_{ij}}(1-\phi_{ckj})^{1-X_{ij}}.
            \]  
      \item Sample $\bm \pi \mid \bm{Y}$ with P\'{o}lya-Gamma augmentation. Let $m^{(s, t, a)} = \logit(\pi^{(s, t, a)})$ and rewrite the model as $\bm{m} = \bm{P} \bm{\eta}$, where $\bm{P}$ is the known design matrix and
   \begin{align*}
   \bm{\eta} &= \begin{pmatrix}
    \mu,\alpha^{\MALE},\alpha_1^{\TIME},...,\alpha_T^{\TIME}, \alpha_1^{\AGE},...,\alpha_A^{\AGE},
    \epsilon_{111},...,\epsilon_{11A}, \epsilon_{121}, ..., \epsilon_{12A},..., \epsilon_{2TA}
    \end{pmatrix}^T,
    \\ 
    \bm{m} &= \begin{pmatrix}
    m^{(1, 1,1)}, ..., m^{(1, 1,A)}, m^{(1, 2,1)}, ..., m^{(1, 2,A)}, ..., m^{(2, T,A)}
    \end{pmatrix}^T.
    \end{align*}
    It suffices to sample $\bm\eta \mid \bm Y$. The prior distribution of $\bm\eta$ is $N(\bm 0, \bOmega^{-1})$, where the prior precision matrix $\bOmega = \mbox{bdiag}(\frac{1}{100}\bI_2, \bOmega_{1}, \bOmega_{2}, \bOmega_{3})$ is block diagonal with $\bOmega_{1}, \bOmega_{2}$ and $\bOmega_{3}$ being the corresponding prior precision matrices of the latent time effect, age effect, and independent noise terms.

Denote $\bm z =  (z^{(1, 1, 1)}, ..., z^{(1, 1, A)}, z^{(1, 2, 1)}, ..., z^{(1, 2, A)}, ..., z^{(2, T, A)})$ with
 \[z^{(s, t,a)} = \sum_{i=1}^{n_{sta}} Y_i \mid m^{(s[i], t[i], a[i])} \sim Bin(n_{sta}, \logit^{-1}(m^{(s[i],t[i], a[i])}))\]
  where $n_{sta}$ denotes the sample size under sex $s$, time period $t$ and age group $a$. We apply P\'{o}lya-Gamma augmentation and sample $\bm\eta$ by introducing latent random variables $\bomega = (\bomega_1, ..., \bomega_{2TA})$: 
  \begin{align*}
    \omega_{l} \mid \bm{\eta} &\sim PG(n_l, \bm{P_l}^{T} \bm{\eta}) \;\;\;\; \mbox{for $l = 1, ..., 2TA$,}
  \\
  \bm{\eta} \mid \bm{\omega}, \bm{z} &\sim MVN(\bm{M}, \bm{V})
  \end{align*}
  where 
  $
    % \bm{M} &= \bm{V}(\bm{\Sigma}^{-1}\bm{\mu} + \bm{P}^{T}\bm{\kappa}) \\
    \bm{M} = \bm{V} \bm{P}^{T}(\bm{z} - \bm{n}/2)
  $
  and 
  $
    \bm{V} = (\bm{\Sigma}^{-1} + \bm{P}^{T} \mbox{diag}(\bomega) \bm{P})^{-1}
  $.

    \item Sample $\phi_{ckj} \mid \bm{Y}, \bm{X}, \bm{Z}$,  for $c = 0,1$, $k = 1, ...,K$, $j = 1, ..., p$,   with 
        \[
        \phi_{ckj} \mid \bm{Y}, \bm{X}, \bm{Z} \sim \mbox{Beta}\left(a_\phi + \sum_{i=1}^{n} \bOne_{\{Y_i=c, Z_i=k, X_{ij}=1\}}, b_\phi + \sum_{i=1}^{n} \bOne_{\{Y_i=c, Z_i=k, X_{ij}=0\}}\right)
        \]
        
    \item Sample $V_{ck}^{(s,t,a)} \mid \bm{Z} ,\bm{Y}, \omega^{(s,t,a)}$,  for $c = 0, 1$, $k = 1, ...,K-1$, $s = 1, 2$, $t = 1, ..., T$,  $a = 1, ..., A$ with 
        \[
        V_{ck}^{(s,t,a)}  \mid  \bm{Z},\bm{Y}, \bm{\omega} \sim \mbox{Beta}\left(1 + \sum_{i=1}^{n_{sta}}  \bOne_{\{ Z_i=k, Y_i = c\}}, \omega_{c}^{(s,t,a)} + \sum_{i=1}^{n_{sta}} \sum_{r=k+1}^{K} \bOne_{\{ Z_i=r, Y_i = c\}}\right)
        \]  
    \item Sample $\omega_{c}^{(s, t,a)} \mid V_{ck}^{(s, t,a)}$,  for $c = 0,1$, $s = 1, 2$, $t = 1, ..., T$,  $a = 1, ..., A$ with 
        \[
        \omega_{c}^{(s,t,a)} \mid V_{ck}^{(s,t,a)} \sim \mbox{Gamma}\left(a_{\omega}+K-1, b_{\omega}-\log(\prod_{k=1}^{K-1}(1-V_{ck}^{(s,t,a)}))\right)
        \]

    \item  Sample $\sigma_{\TIME}^2 \mid \bm{\alpha}^{\TIME}$, $\sigma_{\AGE}^2 \mid \bm{\alpha}^{\AGE}$ with\\
        For the Random Walk model,  
        \[\sigma_{\TIME}^2 \mid \bm{\alpha}^{\TIME}\sim \mbox{Inv-Gamma}
\left(\frac{T-1}{2}+0.5, \frac{\sum_{t=2}^T (\alpha_t^{\TIME} - \alpha_{t-1}^{\TIME})^2}{2} + 0.0009\right)
        \]
        \[\sigma_{\AGE}^2 \mid \bm{\alpha}^{\AGE} \sim \mbox{Inv-Gamma}
\left(\frac{A-1}{2}+0.5, \frac{\sum_{a=2}^A (\alpha_a^{\AGE} - \alpha_{a-1}^{\AGE})^2}{2} + 0.0009\right)
        \]
            For the Independent model, 
        \[\sigma_{\TIME}^2 \mid \bm{\alpha}^{\TIME}\sim \mbox{Inv-Gamma}
\left(\frac{T}{2}+0.5, \frac{\sum_{t=1}^T (\alpha_t^{\TIME})^2}{2} + 0.0015\right)
        \]
        \[\sigma_{\AGE}^2 \mid \bm{\alpha}^{\AGE} \sim \mbox{Inv-Gamma}
\left(\frac{A}{2}+0.5, \frac{\sum_{a=1}^A (\alpha_a^{\AGE})^2}{2} + 0.0015\right)
        \]
  
  \item Sample $\sigma_{\epsilon}^2 \mid \bm{\epsilon}$  with
        \[\sigma_{\epsilon}^2 \mid \bm{\epsilon} \sim \mbox{Inv-Gamma}
\left(\frac{2\times T\times A}{2}+0.5, \frac{\sum_{s=1}^2 \sum_{t=1}^T \sum_{a=1}^A (\epsilon_{sta})^2}{2} + 0.5\right)
        \]
\end{enumerate}

\section{Numerical results}\label{sec:result}

\subsection{Simulated data}\label{sec:simulation}
\begin{figure}[!tb]
    \centering
    \includegraphics[width = 0.8\textwidth]{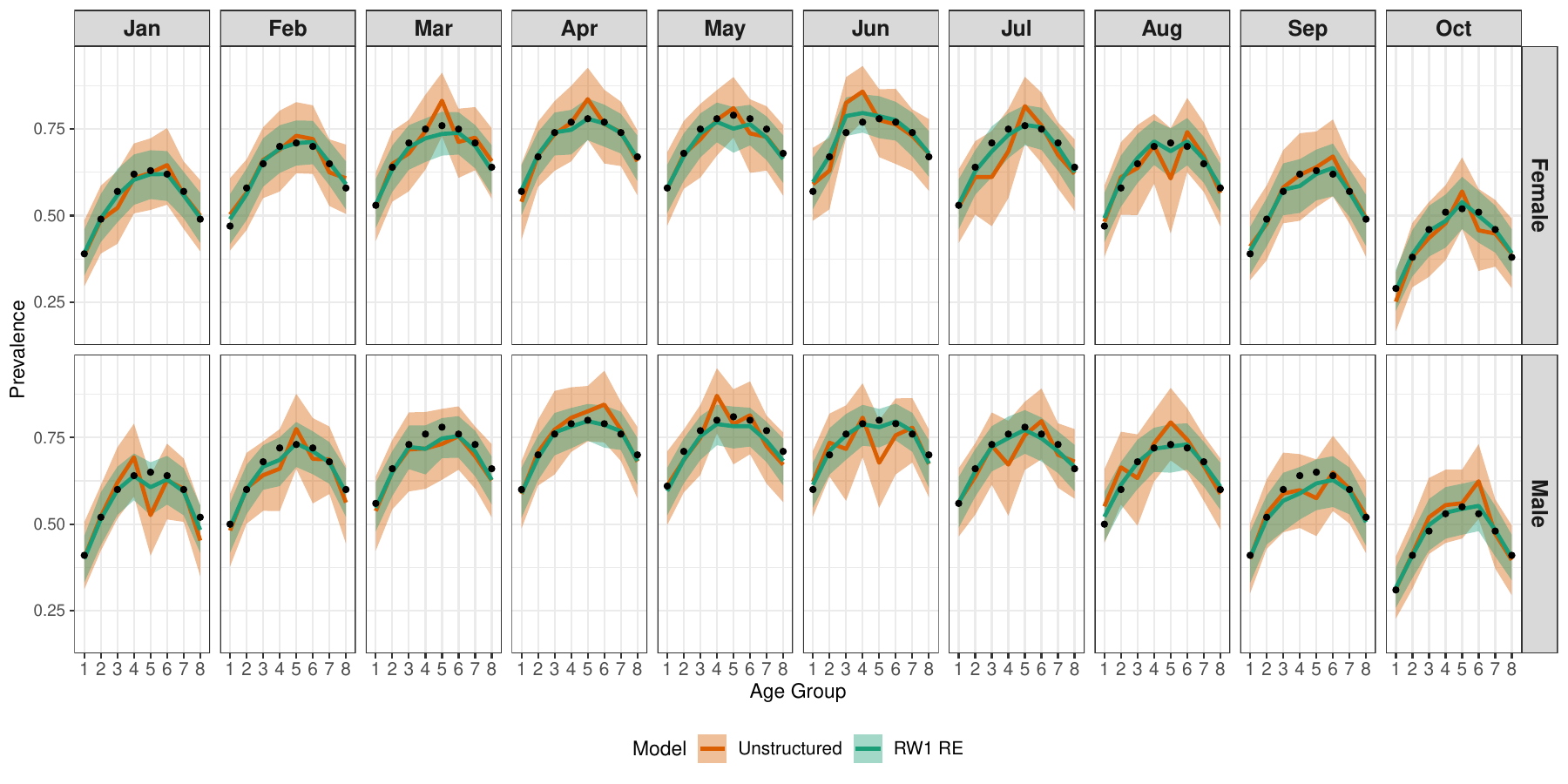}
    \caption{Posterior mean and 95\% credible intervals of the estimated CSMF in one simulated dataset for different age groups, sex, and months. The true CSMF is indicated by the black dots.}
    \label{fig:sim_1_pos_2}
\end{figure}

We start with fully synthetic data simulation. To mimic the true CSMFs across time and age groups, we generate the true $\pi_{sta}$ with the additive polynomial trend in both time and age dimensions. The simulated CSMFs are shown as the black dots in Figure \ref{fig:sim_1_pos_2}. The cause of death $Y$ and symptoms $X$ are generated according to the latent class model described in Section \ref{sec:model} with $K = 10$. In the simulation, we let $p = 10$, $T = 10$, and $A = 8$. The sample size for each sub-population is balanced with $n_{sta} = 100$. We generate $50$ synthetic datasets and fit all models with $K = 10$. 

We consider the following verification mechanism, where the stratification variables include age of the deceased and the time period of death, i.e., $D_i = (A_i, T_i)$. We let
\begin{equation*}   
    p(L_i \mid X_i = \bx, A_i = a, T_i = t, Y_i = y) = \mbox{logit}^{-1}(a^{\TIME}_{t} + a^{\AGE}_{a} + \bm{b}_{ta}^T \bx + c_1 y + c_2 (1-y)).
\end{equation*}
We consider two types of verification mechanisms in the simulation study. 
\begin{itemize}
 \item \textbf{Simulation case (i)}: The verification process depends on observed symptoms $X$ and covariates $D$. We let $c_1 = c_2 = 0$. Assumption \ref{ass:strat} is satisfied in this case.
 \item \textbf{Simulation case (ii)}: The verification process depends on observed symptoms $X$,  covariates $D$, and also the unobserved cause of death $Y$. We let $c_1 \sim \mbox{Unif}(-0.4, 0)$, $c_2 = -c_1$. In this case, the verification process is not conditionally ignorable, and CSMF estimation can be biased. This situation can arise in practice when the verification process depends on variables associated with the cause of death but is unavailable to the data analysts.
\end{itemize}

For the choice of the coefficients in the verification mechanism, we consider the more realistic situations where we under-sample time periods with more deaths and over-sample time periods with fewer deaths. We also over-sample the first two and last two age groups to reflect situations where the mortality patterns of the younger and elderly groups are of greater interest. For the other symptoms, we allow them to be weakly associated with the selection probability in a time-varying fashion. Specifically, we let 
\begin{align*}
    \bm{a}^{\TIME} &= [1.2, 0.1, ..., 0.1, 1.2],
    \\ 
    \bm{a}^{\AGE} &= [0.4, 0.4, -1.6, ..., -1.6, 0.4, 0.4],
    \\
    b_{tj} &= 0.1\bOne_{\{j \in \mathcal{S}_{t}\}}, \quad j = 1, ..., p. 
\end{align*} where $\mathcal{S}_{t} \subset\{1, ..., p\}$ is a set of three randomly chosen indices for each time-level stratum. For the first and last two age groups, this leads to about $60\%$ of deaths being verified in the middle of eight time periods and about $85\%$ verified in the first and last time periods. For the other four age groups, this leads to about $20\%$ of deaths being verified in the middle of eight time periods, and about $40\%$ verified in the first and last time periods.

%  \begin{figure}[!tb]
%     \centering
%     \includegraphics[width = 0.9\textwidth]{figures/simulated_prevalence.pdf}
%     \caption{Simulated prevalence across different age and sex strata over time.}
%     \label{fig:sim_true_prevalence}
% \end{figure}

%\red{Summarize result: I think we don't need to add the full posterior curve comparisons, since it takes a lot of space and is repetitive. How about only show something like simulation\_1\_posterior2.pdf, where you compare unstructured baseline with a set smooth estimates. We should focus on accuracy measures, so bias plot is probably what we want, but need to think how to summarize the different ways (aggregated or not) to show the two set of results (with/without cY). Does CRPS tell any useful stories here?}

In addition to the three models discussed in Section \ref{sec:structure}, to assess the benefits of the proposed models, we consider the following alternative models:

\begin{enumerate}
\item \textbf{Time-only model}: We include sex and age group as binary dummy variables in $X$, and remove the sex- and age-effect from the model for $\bpi$. This leads to a simplified model that is stratified by time. This model represents the scenario where the time-varying CSMF is of interest, but the stratification by sex and age is not accounted for in the prior for $\bpi$. This model ignores the heterogeneity of symptom distributions across different age groups and sexes.  

\item \textbf{Unstructured baseline model}: We use independent priors for $\bpi$ without borrowing information. That is, we let $\pi_{sta} \overset{iid}{\sim} \mbox{Beta}(1, 1)$. This model is a simpler alternative to the structured priors proposed in Section \ref{sec:structure}.

\item \textbf{Unstratified baseline model}: In this model, we simply let $\pi_{sta} = \pi_0 \sim \mbox{Beta}(1, 1)$ for all strata. As we have discussed in Section \ref{sec:stratification}, this unstratified model can lead to biased CSMF estimates if the verification mechanism changes over variables not accounted for by the model. This baseline is only relevant when the target of inference is the overall CSMF.

\end{enumerate}

It should be noted that in terms of modeling $\bpi$, the unstratified baseline model is the default approach in the VA literature and routine analysis when only the overall CSMF is of interest, and the unstructured baseline model is the standard approach when multiple sub-populations are of interest \citep{mccormick2016probabilistic, li2023openva}. In our study, both baseline models adopt the latent class model framework in modeling $p(X\mid Y)$. Thus, they are already more flexible than the VA models currently used in practice, which assume conditional independence of symptoms \citep{mccormick2016probabilistic, byass2019integrated}. Given the extensive evidence that the latent class model framework improves from models making the conditional independent assumption \citep{li2021bayesian, wu2021tree}, we do not further compare our methods to simpler models.

We first illustrate the bias induced by not accounting for the verification process discussed in Section \ref{sec:stratification}. Here we compare the three stratified models with structured priors, the unstructured baseline model, and the unstratified baseline model.  
% This baseline corresponds to the naive adoption of VA cause-of-death assignment algorithms without considering the time-varying verification probability. 
Since the unstratified model aims to estimate the overall CSMF in the population, we aggregate the estimated stratum-specific CSMFs in the other four models to the population-level CSMF by
$$\hat{\pi} = \frac{1}{n} \sum_{s = 1}^2 \sum_{t = 1}^T \sum_{a = 1}^A  n_{sta} \hat{\pi}^{(s, t, a)}.$$

\begin{figure}[!tb]
    \centering
    \includegraphics[width = \textwidth]{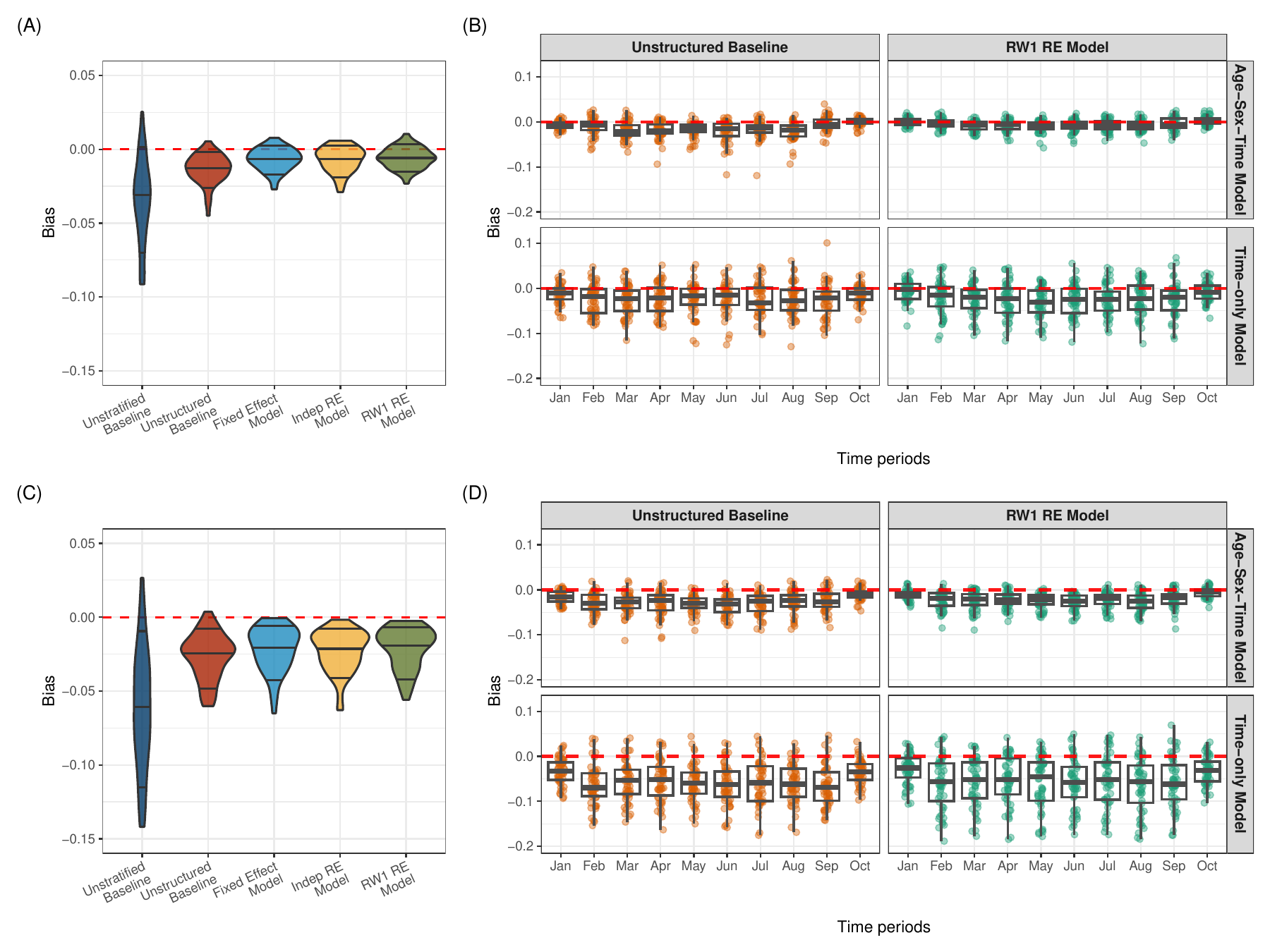}
    \caption{Bias of posterior mean CSMF estimates over $50$ simulated datasets. The top row corresponds to simulation case (i), and the bottom row corresponds to simulation case (ii). Panels (A) and (C) evaluate the bias for population-level overall CSMF estimates, and panels (B) and (D) evaluate the bias for time-varying overall CSMF estimates.}
    \label{fig:sim_box_agg_to_time}
\end{figure}

We take the posterior mean as our point estimate and evaluate the bias compared to the true CSMF, i.e., $\hat \pi - \pi$.
Panel (a) of Figure \ref{fig:sim_box_agg_to_time} shows that in case (i), the unstratified model leads to the largest bias across the simulated datasets. The three models with structured prior perform similarly and achieve much smaller bias compared to the unstructured model. Panel (c) of Figure \ref{fig:sim_box_agg_to_time} shows similar patterns for case (ii). In this case, the proposed models partially account for the verification process and are more robust than the unstratified model even when the missing at random assumption is violated.  

Next, we consider the bias from ignoring the heterogeneity of symptom distributions across strata, discussed in Section \ref{sec:latent}. We consider the comparison between the models stratified by sex, age, and time, and the time-only model. For the fully stratified model, we obtain the time-varying overall CSMF using a similar aggregation step as before, 
$$
\hat{\pi}^{(t)} = \frac{\sum_{s = 1}^2 \sum_{a = 1}^A n_{sta} \hat{\pi}^{(s, t, a)}}{\sum_{s = 1}^2  \sum_{a = 1}^A n_{sta}}.
$$
Panels (b) and (d) of Figure \ref{fig:sim_box_agg_to_time} compare the bias of $\hat{\pi}^{(t)}$ based on the model with partial and full stratification, under the unstructured baseline and the random walk model. The fixed effect and models also show similar patterns. It can be observed that the fully stratified models are able to achieve consistently lower bias in both cases, due to their flexibility to incorporate stratum-specific symptom distribution.

Finally, we evaluate the effect of the structured prior on the stratum-specific CSMF estimates. Figure \ref{fig:sim_1_pos_2} illustrates the posterior means and 95\% credible intervals for the CSMF estimates derived from a single synthetic dataset in case (i). The unstructured baseline model yields CSMF estimates that exhibit notably higher variability over time with larger uncertainty. In contrast, the random walk model yields smoother estimates with narrower credible intervals and generally captures the actual CSMF more accurately.

\subsection{Brazil COVID-19 analysis} \label{sec:brazil}
\begin{figure}[!tb]
    \centering
    \includegraphics[width = \textwidth]{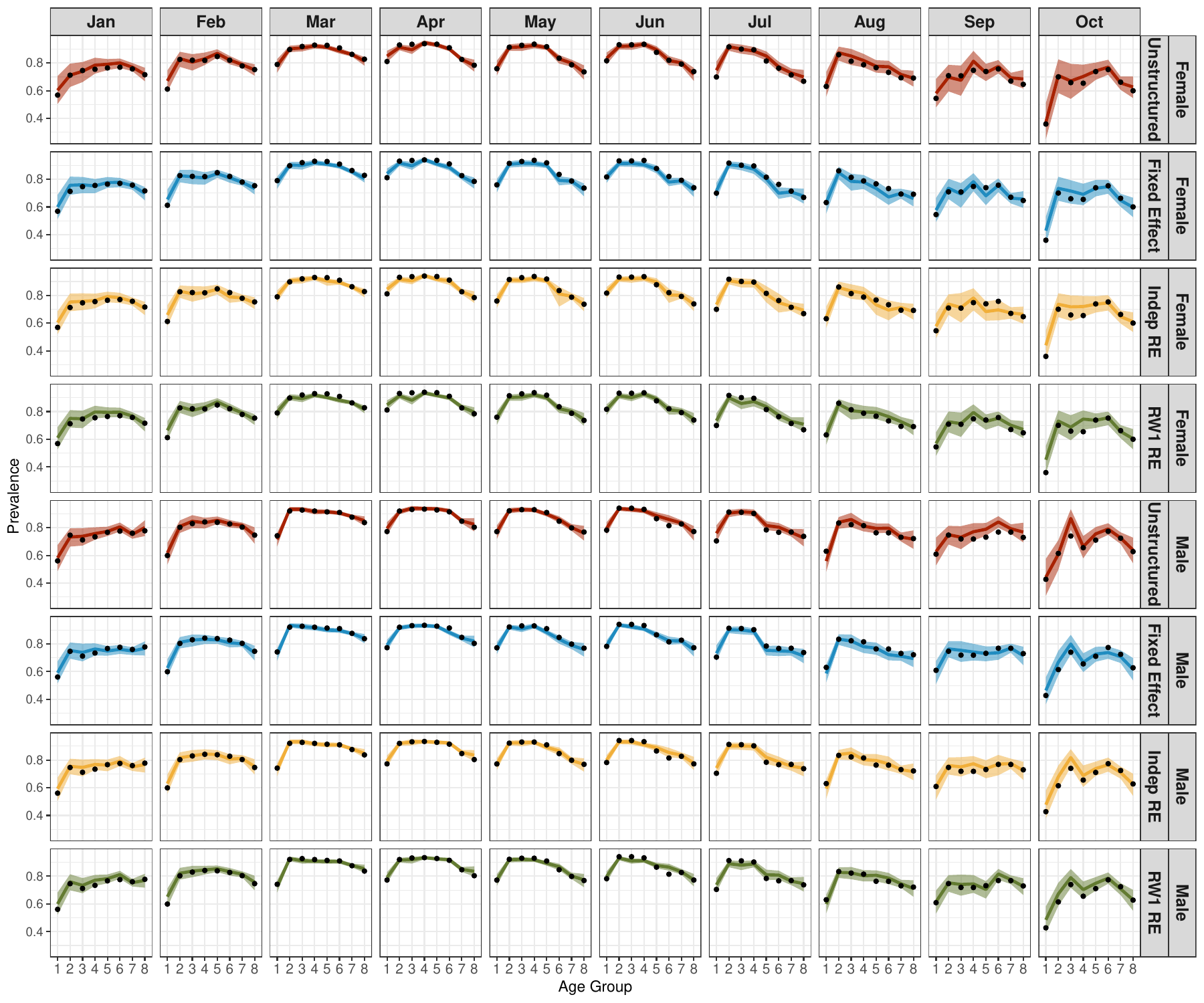}
    \caption{Posterior mean and 95\% credible intervals of the estimated CSMF using the four models for different age groups, sex, and months, based on one synthetic dataset resampled from the COVID-19 surveillance data from Brazil. The true CSMFs are indicated by the black dots.  Age group 1 corresponds to 0--30 years old. Age groups 2 to 7 correspond to 10-year intervals from 30--40 to 80--90. Age group 8 corresponds to over 90 years old.
}
    \label{fig:post_inf_prev}
\end{figure}

We now evaluate our methods using the Brazil COVID-19 surveillance dataset described in Section \ref{sec:data}. We consider the case where $D_i = (S_i, A_i, T_i)$, that is the data are stratified by sex, age, and time periods. To create realistic replications, we generated $100$ semi-synthetic datasets through resampling. In each resampled dataset, we randomly sample $50\%$ observations within each sex, month, and age group, while keeping the proportion of COVID-19 related deaths the same as the CSMFs in the full population. We assume a verification mechanism that depends on time, age, and a random subset of symptoms, with the same coefficient as described in the simulation study,
\begin{equation*}
    p(L_i \mid X_i = \bx, A_i = a, T_i = t) = \mbox{logit}^{-1}(a^{\TIME}_{t} + a^{\AGE}_{a} + \bm{b}_{ta}^T \bx).
\end{equation*}
We fit all models with $K = 10$. We run the MCMC for $8000$ iterations with $3000$ iterations as burn-in. 
Figure \ref{fig:post_inf_prev} shows the posterior mean and 95\% credible interval of the stratum-specific CSMF in one resampled dataset, estimated by the four stratified models. Stronger effects of smoothing can be observed in the two random effect models, especially in the last two time periods when the total number of deaths is smaller.

% Figure \ref{fig:bias_boxplot} provides a comprehensive summary of the bias in the estimated prevalence across 50 synthetic datasets, considering four different prior choices. The baseline model exhibits larger absolute bias and demonstrates highly unstable point estimates of prevalence. This indicates that without the inclusion of information-sharing priors, the baseline model struggles to accurately estimate prevalence in a consistent manner.

% On the other hand, the three models that incorporate priors that borrow information across strata show similar biases, which are noticeably smaller compared to the baseline model. This implies that the inclusion of structured priors leads to clear improvements in prevalence estimation. The effect of the structured prior becomes even more evident when examining the estimates for individual synthetic datasets.

\begin{figure}[!tb]
    \centering
    \includegraphics[width = 0.7\textwidth]{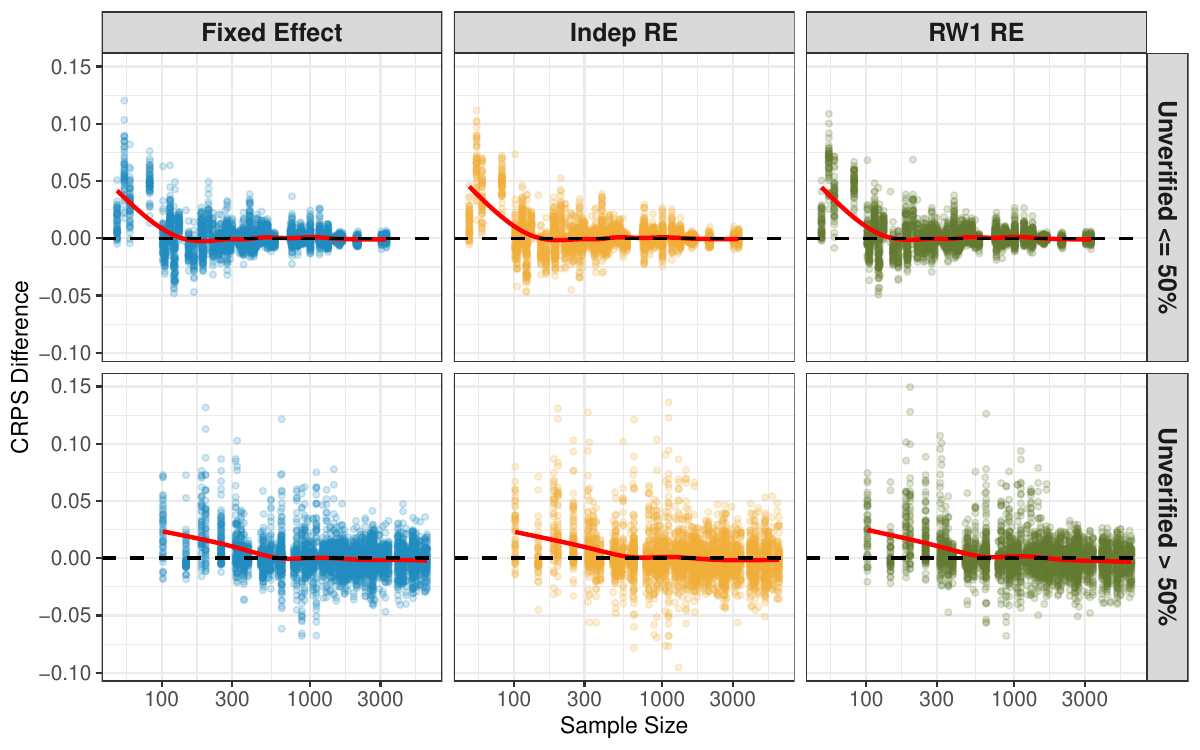}
    \caption{Improvement in CRPS compared to the unstructured baseline model for the three structured models, in terms of difference in CRPS. Each dot represents one sub-population in one simulated dataset, arranged by the sample size of the sub-population and the proportion of unverified labels, i.e., the proportion of deaths without a cause, in the sub-population. The red line is the smoothed conditional mean for the points.}
    \label{fig:CRPS_scatterplot}
\end{figure}

To further compare the predictive performance of the four models, we compute the Continuous Ranked Probability Score (CRPS) \citep{gneiting2007strictly} for all four models. CRPS is an extension of the squared error loss that takes into account the full predictive distribution instead of only a point estimate. For a probabilistic prediction distribution $F$, CRPS is defined as follows,
\[CRPS(F, x) = E_F|X - x| - \frac{1}{2} E_F|X - X'|.
\]
where $X \sim F$ and $X' \sim F$ are two independent random variables, and $x$ is the true CSMF. CRPS is positive, where values closer to 0 indicate better prediction. We compute the CRPS difference between the baseline and each of the three structured models. Positive differences indicate that the structured models outperform the baseline. Figure \ref{fig:CRPS_scatterplot} shows the CRPS differences for the three models, compared to the unstructured baseline model, across all sub-populations in the $50$ datasets. As expected, we can observe improved CRPS from all three structured models when the sample size is small and the proportion of unverified deaths is high.

\begin{figure}[!tb]
    \centering
        \includegraphics[width = 0.6\textwidth]{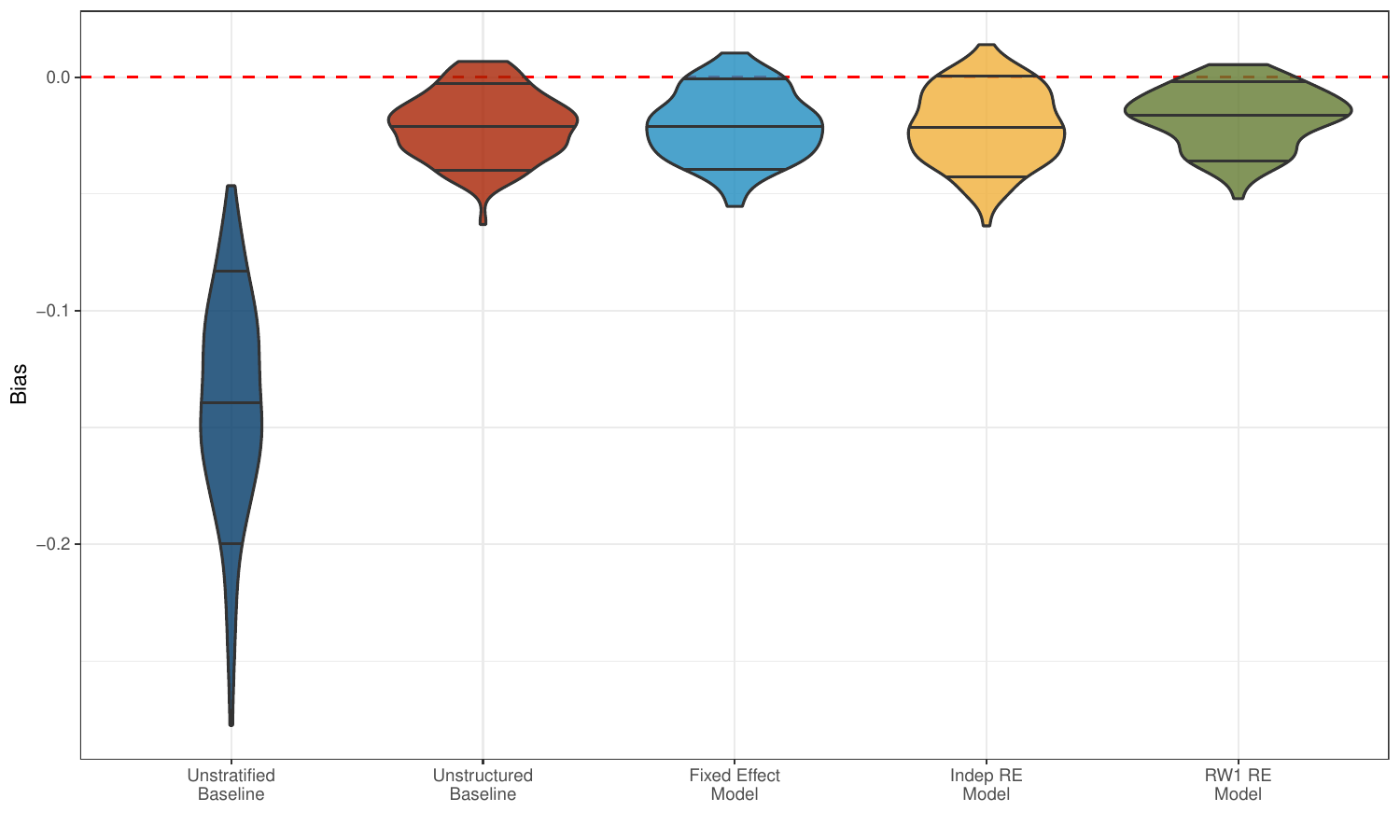}
    \caption{Distribution of bias of overall CSMF estimation over $100$ synthetic datasets resampled from the COVID-19 surveillance data from Brazil, under the population-level model and the four stratified models.}
    \label{fig:box_agg_to_all}
\end{figure}

\begin{figure}[!tb]
    \centering
        \includegraphics[width = 0.8\textwidth]{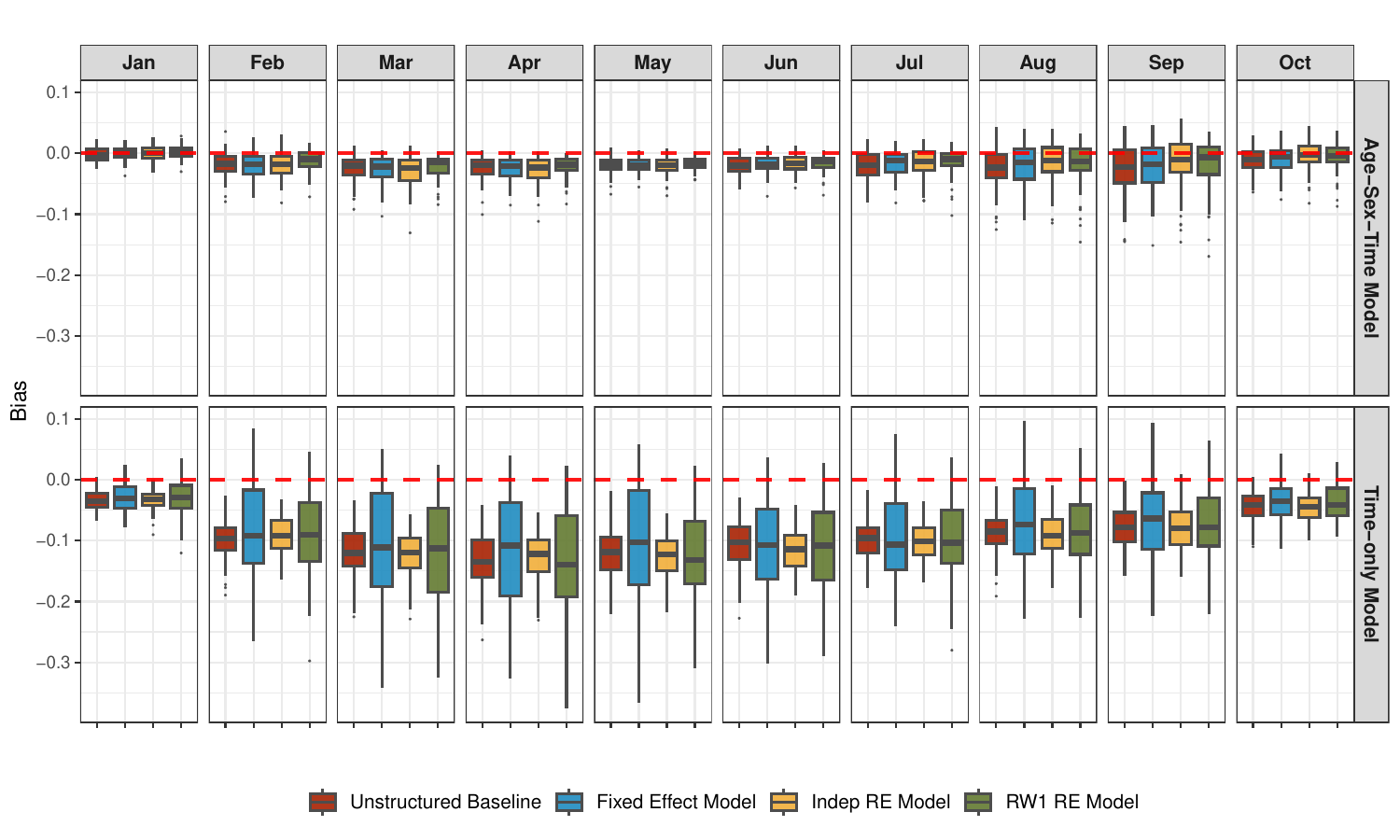}
    \caption{Bias of time-varying CSMF estimation over $100$ synthetic datasets resampled from the COVID-19 surveillance data from Brazil, under the four models stratified by sex, age, and time (top row) and the models stratified by time only (bottom row).}
    \label{fig:box_agg_to_time}
\end{figure}

% \subsubsection{Aggregated prevalence estimation}
Figure \ref{fig:box_agg_to_all} shows the similar pattern of bias as illustrated in the simulation study. The unstructured baseline model results in the largest bias for the population-level overall CSMF, due to not accounting for the time-varying verification mechanism. 
Figure \ref{fig:box_agg_to_time} shows that for all models, accounting for the heterogeneity of symptom distributions over sex and age also greatly improves CSMF estimation.

\begin{figure}[!tb]
    \centering
        \includegraphics[width = \textwidth]{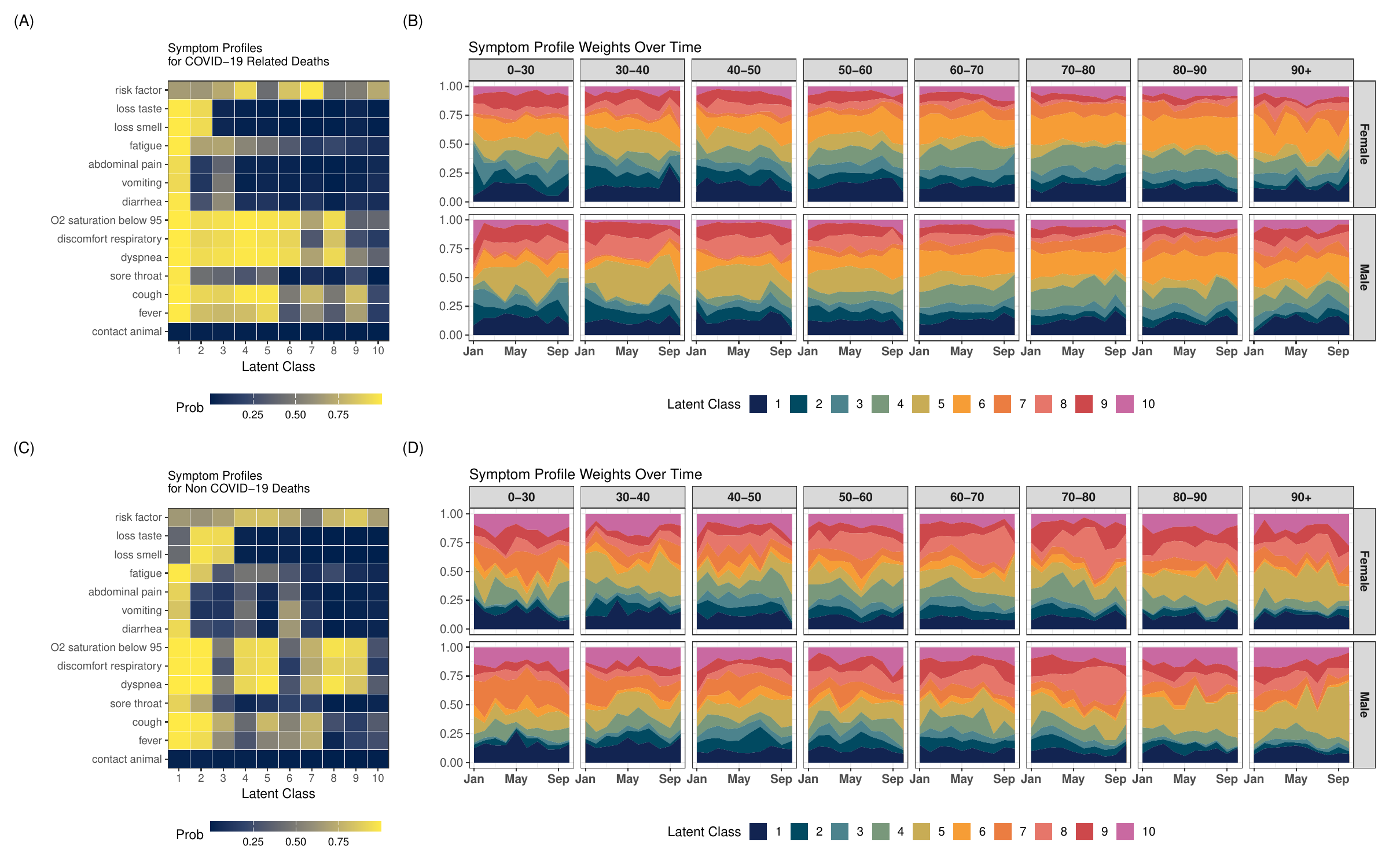}
    \caption{Posterior mean of latent parameters estimated in one synthetic dataset resampled from the COVID-19 surveillance data from Brazil. Panel (a): symptom profiles $\bphi$ for deaths related to COVID-19, ordered by the expected number of symptoms. Each column corresponds to one latent class, and the color correspond to the probability of observing the symptom within the corresponding latent class. Panel (b): latent class weights $\blambda$ over sex, age, and time for deaths related to COVID-19. Panel (c) and (d) are the same parameters for deaths not related to COVID-19.}
    \label{fig:heatmap_latent_weights}
\end{figure}

% \subsubsection{Latent class indications across strata} 
Finally, the proposed latent class model framework also allows us to gain more insight into how the data distribution shifts over time by examining the estimated latent parameters. Figure \ref{fig:heatmap_latent_weights} visualizes the heatmap of latent symptom profiles and the corresponding weights across different strata under the random walk model in one synthetic dataset. The latent classes are ordered by the expected number of symptoms under the symptom profiles of each class. We can observe distinct clusters of symptoms, such as fever and cough; dyspnea, respiratory discomfort, and low $O_2$ saturation; loss of taste and loss of smell, that are more likely to occur in the same latent symptom profiles, which seem to correspond to symptoms with shared mechanisms. The estimated weights $\blambda$ illustrate the changes in abundance of these latent classes over age, sex, and time. For example, among the deaths related to COVID-19, the weights of the fifth latent class are much larger in male deaths under 60 years old, compared to other demographic groups, whereas the sixth latent class has larger weights among deaths above 60 years old. Both symptom profiles include high probabilities of dyspnea, respiratory discomfort, and low $O_2$ saturation. The main difference in these two latent symptom profiles is that the former has a higher probability of observing fever and cough, whereas the latter has a higher probability of having at least one risk factor. These patterns are consistent with the explorative analysis in Figure \ref{fig:symptoms_dist}.

\section{Discussion}\label{sec:discuss} 
In this paper, we introduce a hierarchical latent class model framework for modeling partially verified VA data and estimating the sub-population CSMFs of a target cause of death. We describe conditions under which valid inference of the CSMF can be made and propose the novel use of structured priors to borrow information across sub-populations. We demonstrate that our model is able to avoid the bias induced by informative sampling of reference deaths, and the structured priors can improve the robustness and interpretability of the estimated CSMF. 

Our work also highlights the need to carefully consider how training data are collected when deploying predictive models in high-stakes health surveillance tasks. In practice, as long as deaths are selected for verification according to a well-defined study design and protocol, the missing at random assumption can be made plausible by including all relevant factors that influence the verification mechanism in the model. Therefore, careful documentation and consideration of the study design are essential. Failures to account for the data collection process can lead to severely biased results, even with sophisticated models, an aspect of particular relevance as the use of machine learning and artificial intelligence becomes increasingly popular in population health research.

There are several limitations of the proposed model. First, in the context of classifying COVID-19 deaths, information on PCR or antigen tests provides a key piece of information. However, the availability of tests needs to be taken into account and we generally should not treat it as missing at random. More generally, an important direction of future research is to deal with non-random missing indicators in both $X_E$ and $X_C$. 
Second, we treat the latent class distribution to be independent a priori in the current model. When sample sizes are small, further incorporating structured priors on the latent class probabilities $\bm{\lambda}$ could further improve the estimation of the indicator distributions and may lead to better overall classification performance.
Finally, existing VA algorithms largely take an overly simplified view on the causal structures among the collected indicators. While we have explored in this paper the role of two indicators, i.e., age and sex, as a stratification variable, more work is needed to further develop a more comprehensive framework to incorporate causal structures among all the collected indicators into the VA models, as such information could be key for proper generalizability of the algorithm across different populations.

\section*{Data Availability Statement}\label{sec:code}

All of the computer codes for this paper are available at \url{https://github.com/YuZoeyZhu/COVID_VA}. The repository also includes a reproducible report demonstrating model fitting using synthetic data. The full Brazil COVID-19 dataset can be shared upon request with the permission of the Brazilian Ministry of Health.

\section*{Acknowledgments}
We would like to thank Dr. Fatima Marinho and Dr. Luiz Fernando Ferraz da Silva for sharing the Brazilian COVID-19 surveillance data and helpful discussion of the results.

ZY and ZRL were supported by grant R03HD110962 from the Eunice Kennedy Shriver National Institute of Child Health and Human Development (NICHD), and in part by the Bill \& Melinda Gates Foundation. The findings and conclusions contained within are those of the authors and do not necessarily reflect positions or policies of the Bill \& Melinda Gates Foundation. 

\clearpage
\bibliographystyle{apalike}
\bibliography{crms}

%USE THE BELOW OPTIONS IN CASE YOU NEED AUTHOR YEAR FORMAT.
%\bibliographystyle{abbrvnat}
%\bibliography{reference}

% %% sample for biography with author's image
% \begin{biography}{{\color{black!20}\rule{77pt}{77pt}}}{\author{Author Name.} This is sample author biography text. The values provided in the optional argument are meant for sample purposes. There is no need to include the width and height of an image in the optional argument for live articles. This is sample author biography text this is sample author biography text this is sample author biography text this is sample author biography text this is sample author biography text this is sample author biography text this is sample author biography text this is sample author biography text.}
% \end{biography}

% %% sample for biography without author's image
% \begin{biography}{}{\author{Author Name.} This is sample author biography text this is sample author biography text this is sample author biography text this is sample author biography text this is sample author biography text this is sample author biography text this is sample author biography text this is sample author biography text.}
% \end{biography}

\end{document}